\documentclass[aps,prd,twocolumn,showpacs,eqsecnum,nofootinbib]{revtex4}
\usepackage[dvips]{epsfig}

\begin{document}

\title{Potential for Supernova Neutrino Detection in MiniBooNE}

\author{
\mbox{Matthew K. Sharp$^{1}$}, 
\mbox{John F. Beacom$^{2}$},
\mbox{Joseph A. Formaggio$^{1}$}}

\altaffiliation{Present address: Center for Experimental Nuclear
Physics and Astrophysics, University of Washington, Seattle, WA 98195, USA}

\affiliation{
\mbox{$^1$ Physics Department, Columbia University, New York, New York 
10027, USA} 
\mbox{$^2$ NASA/Fermilab Astrophysics Center, Fermi National Accelerator 
Laboratory, Batavia, Illinois 60510-0500, USA}
\\
{\tt msharp@phys.columbia.edu}, 
{\tt beacom@fnal.gov}, 
{\tt josephf@u.washington.edu}}

\date{3 May 2002}

\begin{abstract}
The MiniBooNE detector at Fermilab is designed to search for $\nu_\mu
\rightarrow \nu_e$ oscillation appearance at $E_\nu \sim 1 {\rm\ GeV}$
and to make a decisive test of the LSND signal.  The main detector
(inside a veto shield) is a spherical volume containing 0.680 ktons of
mineral oil.  This inner volume, viewed by 1280 phototubes, is
primarily a \v{C}erenkov medium, as the scintillation yield is low.
The entire detector is under a 3 m earth overburden.  Though the
detector is not optimized for low-energy (tens of MeV) events, and the
cosmic-ray muon rate is high (10 kHz), we show that MiniBooNE can
function as a useful supernova neutrino detector.  Simple
trigger-level cuts can greatly reduce the backgrounds due to
cosmic-ray muons.  For a canonical Galactic supernova at 10 kpc, about
190 supernova $\bar{\nu}_e + p \rightarrow e^+ + n$ events would be
detected.  By adding MiniBooNE to the international network of
supernova detectors, the possibility of a supernova being missed would
be reduced.  Additionally, the paths of the supernova neutrinos
through Earth will be different for MiniBooNE and other detectors,
thus allowing tests of matter-affected mixing effects on the neutrino
signal.
\end{abstract}

\pacs{97.60.Bw, 95.55.Vj        \hspace{5cm} FERMILAB-Pub-02/073-A}

\maketitle


\section{Introduction}

As is well-known, about two dozen neutrinos in total were detected
from SN1987A in the Kamiokande II, IMB, and Baksan
detectors~\cite{SN1987A}.  Even these very limited observations,
despite some of their puzzling features, did provide a basic
confirmation of the core-collapse supernova mechanism as well as
interesting limits on the properties of neutrinos~\cite{raffeltbook}.
The Galactic supernova rate is about $(3 \pm 1)$/century (most would
be obscured optically by dust)~\cite{SNrate}, so it is very important
that a supernova neutrino signal not be missed because of detectors
being down for upgrades or calibrations.  This can be accomplished by
having as many independent supernova neutrino detectors as possible.
Since different detectors use different targets and techniques, having
results from several detectors is also very useful for making
cross-checks of the data and theory.  Additionally, the neutrino paths
through Earth will be different, and matter-affected mixing effects on
the signal can be significant (see, e.g., Ref.~\cite{lunardini}).

The supernova neutrino detection capabilities of various present or
near-term detectors are documented elsewhere: Super-Kamiokande
(SK)~\cite{SNsk}, the Sudbury Neutrino Observatory (SNO)~\cite{SNsno},
Borexino~\cite{SNborexino}, KamLAND~\cite{SNkamland}, the Large Volume
Detector (LVD)~\cite{SNlvd}, and AMANDA~\cite{SNamanda}.  SK, once
repaired, would expect about $10^4$ identified supernova events.  The
others would expect between a few and several hundred identified
events.  (The number of identified events in AMANDA is more difficult
to quantify since the supernova is seen only as a statistically
significant increase in the noise rate).  The yields are expected to
be larger than from SN1987A in part because the assumed distance is
smaller.  SN1987A was at a distance of about 50 kpc, in the Large
Magellanic Cloud, a small companion of the Milky Way Galaxy.  The next
supernova will more likely be in our Galaxy proper, and
conventionally, a distance of 10 kpc is assumed, approximately the
median distance of Galactic stars from Earth.  In the case of SK, it
is approximately 16 times larger than its predecessor Kamiokande II.

The MiniBooNE detector at Fermilab is designed to search for $\nu_\mu
\rightarrow \nu_e$ oscillation appearance, using a beam of $\sim 1
{\rm\ GeV}$ $\nu_\mu$ produced by $\pi^+/K^+$ decay in flight.  These
mesons are produced when a proton beam from the Fermilab Booster hits
a beryllium target about 500 m away from the detector.  The mesons are
focused by a magnetic horn system that will allow charge selection and
hence running with antineutrinos instead of neutrinos.  The beam will
operate with the very low duty cycle of 5 Hz of 1.6 $\mu s$ spills, so
only modest shielding from cosmic-ray muons is required.  This is
provided by a 3 m earth overburden, which nearly eliminates the
hadronic component of the cosmic rays (the hadronic interaction length
is about 1 m water equivalent).  The MiniBooNE experiment will
decisively confirm or refute the LSND~\cite{lsnd} neutrino oscillation
signal; full operations begin in Summer 2002.

We briefly review the basic characteristics of the MiniBooNE detector.
A more complete description can be found in Ref.~\cite{miniboone}.
The detector is a 6.1 m radius steel sphere, filled with mineral oil.
The oil has density 0.85 g/cm$^3$ and its chemical composition is
C$_{n}$H$_{2n+2}$, with $n \simeq 30$.  At 5.75 m radius, there is a
phototube support structure that optically isolates the inner volume
from a veto region.  The veto region is painted white to maximize
light-gathering efficiency (\v{C}erenkov imaging will thus not be
possible); it is viewed by 241 phototubes, and is expected to have
greater than $99 \%$ efficiency for detecting cosmic-ray muons
crossing the veto region once.  The inner volume, containing 0.680
ktons of oil, is designed as an imaging \v{C}erenkov detector viewed
by 1280 phototubes providing $10\%$ photocathode coverage.  For the
main oscillation experiment, a fiducial volume will be defined inside
5 m radius, and containing 0.445 ktons of oil.  Though no
scintillating compounds have been added to the oil, there is still
some light from scintillation.  We assume 4 photoelectrons per MeV,
with a $3:1$ ratio of \v{C}erenkov to scintillation light.  The total
cosmic-ray muon rate in the detector is about 10 kHz, with about 8 kHz
throughgoing and 2 kHz stopping.  These rates were estimated directly
using the known sea-level rates~\cite{RPP}, and are in agreement with
preliminary measurements in the detector.  In the very near future, as
the detector is commissioned and calibrated, the detector properties
will be well-measured, and full Monte Carlo modeling of supernova
neutrino detection will be done.

While the detector is clearly not optimized for detecting supernova
neutrinos, since $\sim 200$ $\bar{\nu}_e + p \rightarrow e^+ + n$
events would be expected, it is worth examining whether it can indeed
be used as a supernova neutrino detector.  In this paper, we show that
with just simple trigger-level cuts, MiniBooNE can efficiently operate
as a supernova neutrino detector without interfering with its main
task of testing the LSND~\cite{lsnd} signal.  This is despite the
likely skepticism to the idea that a surface-level detector could
reduce its cosmic-ray muon backgrounds enough to function as a
supernova neutrino detector.


\section{The Supernova Signal}

In large stars (greater than about $8 M_\odot$), nuclear fusion
reactions begin with protons and eventually proceed through heavier
nuclei until iron is produced.  Since iron is the most tightly-bound
nucleus, the energy generation rate in the core falls as the fraction
of iron increases.  Once the iron core has reached about $1.5
M_\odot$, it can no longer be supported by even electron degeneracy
pressure, and it collapses.  Once nuclear densities are reached, the
core cannot be compressed further, and rebounds, with the subsequent
outgoing shock ejecting the stellar envelope.  In this paper, we
characterize the supernova neutrino signal in a very simple way,
though consistently with numerical supernova models~\cite{SNmodels}.
The change in gravitational binding energy from the stellar core and
the proto-neutron star is about $3 \times 10^{53}$ ergs, about $99\%$
of which is carried off by all flavors of neutrinos and antineutrinos
over about 10 s.  The emission time is much longer than the
light-crossing time of the proto-neutron star because the neutrinos
are trapped and must diffuse out, eventually escaping with
approximately Fermi-Dirac spectra characteristic of the surface of
last scattering.  In the canonical model, $\nu_\mu, \nu_\tau$ and
their antiparticles have a temperature $T \simeq 8$ MeV, $\bar{\nu}_e$
has $T \simeq 5$ MeV, and $\nu_e$ has $T \simeq 3.5$ MeV.  The
temperatures differ from each other because $\bar{\nu}_e$ and $\nu_e$
have charged-current opacities (in addition to the neutral-current
opacities common to all flavors), and because the proto-neutron star
has more neutrons than protons.  It is generally assumed that each of
the six types of neutrino and antineutrino carries away about $1/6$ of
the total binding energy, though this has an uncertainty of at least
$50\%$~\cite{raffeltproc}.

In this paper, we will focus on just the $\bar{\nu}_e$ signal.  The
spectrum shape for the supernova events is given by the product of the
cross section and a Fermi-Dirac distribution, i.e.,
\begin{equation}
\frac{dN}{dE_\nu}
\propto \sigma(E_\nu) \frac{E_\nu^2}{1 + \exp(E_\nu/T)}\,.
\end{equation}
This is the spectrum of neutrinos which interact.  The detection
reaction in MiniBooNE is $\bar{\nu}_e + p \rightarrow e^+ + n$, and
the corresponding positron spectrum is immediately obtained if we
assume that $E_e = E_\nu - 1.3$ MeV (i.e., if neutron recoil is
neglected).  The cross section in this approximation~\cite{invbeta} is
\begin{equation}
\sigma(E_\nu) = 
0.0952 \times (E_\nu - 1.3)^2\,,
\end{equation}
where energies are in MeV and and the cross section is in units of
$10^{-42} {\rm\ cm^2}$.  The full cross section, including the recoil,
weak magnetism, and radiative corrections is given by Vogel and
Beacom~\cite{invbeta}.  For a temperature $T$, the positron spectrum
peaks at about $4T$~\cite{SNsno} (for comparison, the average neutrino
energy before weighting by the cross section is $3.15 T$).

The expected number of events (assuming a hydrogen to carbon ratio
in the detector of $2:1$) is
\begin{eqnarray}
N & = &
11.8
\left[\frac{E_B}{10^{53}{\rm\ erg}}\right]
\left[\frac{1{\rm\ MeV}}{T}\right] \nonumber \\ 
& \times &
\left[\frac{10{\rm\ kpc}}{D}\right]^2
\left[\frac{M_D}{1{\rm\ kton}}\right]
\left[\frac{\langle \sigma \rangle}{10^{-42}{\rm\ cm^2}}\right]\,.
\end{eqnarray}
As noted, we will assume $E_B = 3 \times 10^{53}$ ergs, $T = 5$ MeV,
and $D = 10$ kpc.  We assume that all events within a radius of 5.5 m
can be used, corresponding to 0.595 ktons.  Though the optical barrier
is at 5.75 m radius, the phototubes faces are at 5.5 m radius.  The
positrons have very short range (they lose about 2 MeV/g/cm$^2$) and
are nearly isotropically directed.  For the thermally-averaged cross
section per ${\rm C H}_2$ ``molecule'' (2 protons) we use $\langle
\sigma \rangle = 54 \times 10^{-42} {\rm\ cm^2}$ at $T = 5$ MeV.
Including the corrections of Ref.~\cite{invbeta} would reduce the
thermally-averaged cross section by about $20\%$; in the present
study, these corrections may be neglected.

Thus the total yield from $\bar{\nu}_e + p \rightarrow e^{+} + n$ is
expected to be $N \simeq 230$.  The positrons will be detected in
MiniBooNE by their \v{C}erenkov (and scintillation) light.  The
neutrons will be radiatively captured on protons, but we assume that
the resulting 2.2 MeV gamma rays will not be visible, due to
low-energy radioactivity backgrounds.

For $\bar{\nu}_e + p \rightarrow e^+ + n$, the yield is nearly
proportional to $T$ (since $\langle \sigma \rangle \sim T^2$), and as
noted, the peak of the positron spectrum is about $4 T$.  The true
temperature may be somewhat different (see, e.g.,
Ref.~\cite{raffeltproc,horowitz}), and it may be effectively increased
by mixing with $\bar{\nu}_\mu/\bar{\nu}_\tau$ (see, e.g.,
Ref.~\cite{SNosc}).  We neglect possible distortions in the tail
characterized by a chemical potential, as their effects are minimal
for this cross section~\cite{SNsno}.

The next-most important reaction in the detector will be the
neutral-current nuclear excitation of $^{12}$C, which yields a 15.11
MeV gamma, with $\simeq 30$ events expected (see, e.g.,
~\cite{SNborexino}).  These gammas will Compton-scatter multiple
electrons to a variety of energies.  In the present study, we neglect
these events.  We also neglect the smaller numbers of events from
neutrino-electron scattering and charged-current reactions on
$^{12}$C.

To simulate the energy resolution of the detector, we first consider
the minimum energy resolution that occurs because of the Poisson
statistics of the number of photoelectrons.  For a detector with
$\alpha$ {\it detected} photoelectrons per MeV, the minimum energy
resolution is
\begin{equation}
\delta(E) = \frac{\sqrt{E}}{\sqrt{\alpha}}\,,
\end{equation}
where all energies are in MeV.  Note that $\alpha$ varies from
detector to detector and depends on the number and efficiency of the
phototubes, their distance from the fiducial volume, light absorption,
the fraction of tubes that are multiply hit, etc.  It is therefore
generally a function of position and direction.  In other \v{C}erenkov
detectors, e.g., SK and SNO, all of these effects and more are modeled
in the Monte Carlo, and energy resolution close to the Poisson limit
can be obtained (only about $25\%$ worse).  We assume $\alpha = 4$ for
MiniBooNE (SK and SNO have $\alpha = 6$ and 9, respectively), and that
similar event reconstruction techniques can be employed.  At the
energies of interest, the detector efficiency is taken to be unity.

We conservatively assume that the energy resolution in MiniBooNE will
be about 1.5 times the minimum given by Poisson statistics above.  In
LSND, the energy resolution was about 2.5 times worse than Poisson for
reasons that had to do with the very high light yield due the
scintillating compounds added to their mineral oil.  Because of the
large number of multiply-hit phototubes, energy was estimated by
integrated charge, rather than simply by the number of hit phototubes.
In those phototubes, the charge distribution per photoelectron is very
broad, and has a long tail at high charge. Though the same phototubes
are being used in MiniBooNE, we do not expect to have these problems.
MiniBooNE will not have such a high light yield (the ratio of
\v{C}erenkov light to scintillation light should be $3:1$ instead of
$1:4$, and $\alpha = 4$ instead of $30$), and approximately 300 new
phototubes with better charge resolution have been added.  Therefore,
the energy resolution at low energies in MiniBooNE should be rather
good.


\section{Backgrounds}

We have shown that about 230 $\bar{\nu}_e + p \rightarrow e^+ + n$
events are expected in MiniBooNE from a canonical Galactic supernova
at 10 kpc, and that the positron spectrum peaks at about 20 MeV.  If
these events can be separated from backgrounds, then this is a
respectable yield of events, approximately 10 times more than were
observed in total from SN1987A.  As we show below, the spectrum shape
should be well-measured too.

The key question, of course, is whether these signal events can be
separated from the large cosmic-ray related backgrounds expected in a
surface-level detector (for comparison, SK and SNO are under about 1
and 2 km of rock, respectively).  As noted above, hadronic cosmic rays
will be reduced to a negligible rate by the 3 m earth overburden.  All
of the backgrounds that we consider are related to cosmic-ray muons,
and their total rate through the detector is about 10 kHz (8 kHz
throughgoing, 2 kHz stopping).  How then can we see the $230/10 {\rm\
s} \simeq 20$ Hz supernova signal underneath the 10 kHz muon rate?  In
this Section, we study the background rates in detail and show how
they can be greatly reduced with simple trigger-level cuts.


\subsection{Muon Energy Loss}

We first consider direct energy deposition by muons.  If the veto
shield were perfectly efficient, then any muon in the main detector
volume would be identified by its signal(s) in the veto.  If
throughgoing and stopping muons can be easily distinguished by their
signals in the veto and main detector, then we would only have to
consider possible Michel electrons from muon decays for the 2 kHz of
stopping muons, and not the full muon rate of 10 kHz, thus minimizing
the detector deadtime (this is discussed below).

However, since the muon rate is so high, an appreciable rate (2 kHz
$\times\ 0.01 \simeq$ 20 Hz) of muons can evade one veto layer and
then stop in the detector.  Sea-level muons have average energies of
about 4 GeV, and will lose about 1.6 GeV in the 3 m earth overburden
(assuming 2 MeV/g/cm$^2$ for a minimum-ionizing muon, and that the 3 m
earth overburden is about 8 m water equivalent).  Therefore, a typical
muon might travel about 14 m in oil (density 0.85 g/cm$^3$).  The
spectrum of muon energies is falling only slowly in this energy range,
so a broad distribution of path lengths in the detector is expected.
Therefore, very few muons will lose less than 100 MeV or so, which
would correspond to about 60 cm for a minimum-ionizing muon.
Therefore, direct energy deposits by unvetoed stopped muons will
always be so large as to be easily distinguishable from the supernova
signal.  One might also consider corner-clipping throughgoing muons,
to which similar considerations apply; such muons will also have two
chances to trigger the veto.

Therefore, we will define a muon event as any event in which the
number of hit phototubes in the veto {\bf OR} the main detector is
large.  This is easy to implement as a trigger-level cut, and it solves
the problem of the veto inefficiency.


\subsection{Muon Decays}

Most of the stopped muons in the detector will decay, and the Michel
electrons and positrons from muon decay have an energy spectrum
\begin{equation}
\frac{dN}{dE_e} \propto E_e^2\, (1 - 0.013 E_e)\,,
\end{equation}
where all energies are in MeV and the kinematic endpoint of the
spectrum is 52.8 MeV.  The normalization of the spectrum is set by the
rate of stopped muons, namely 2 kHz.  This is a potentially very
important background, since the event energies are similar to those of
supernova events.

We can dramatically reduce the Michel background by imposing a holdoff
of $15.2\ \mu$s after every muon (the muon lifetime is 2.2 $\mu s$).
During this holdoff period, no data will be taken, which creates a
detector deadtime fraction.  In the ideal case, the holdoff would only
be applied for stopping muons, so that the deadtime fraction would be
$2 {\rm\ kHz} \times 15.2\ \mu{\rm s} = 0.03$, which would be
negligible.  However, because the veto is not perfect, we have to
apply this holdoff after any muon event, as defined above, so the
deadtime fraction will be $10 {\rm\ kHz} \times 15.2\ \mu{\rm s} =
0.15$, which is still small.  Most muon events are throughgoing, and
so will not actually have a Michel decay electron.  If true
throughgoing events can be flagged at the trigger level, then the
deadtime fraction can be reduced.  Similarly, if the positions of true
stopping muons could be determined, then only events nearby in
distance and time would be excluded, instead of making the whole
detector dead (for example, SK uses this technique to avoid large
deadtime).  With the long holdoff of $15.2\ \mu{\rm s}$, we will cut
all but a fraction $10^{-3}$ of Michel decays, so that the true rate
of surviving Michels will be an extremely small 2 Hz in the main
detector volume.  Note that if the holdoff time is reduced, the
deadtime fraction decreases linearly, but the surviving Michel rate
increases exponentially.


\subsection{Beta Decays of $^{12}$B}

Of the 2 kHz of stopped muons, about $44\%$ are
$\mu^-$~\cite{muratio}, of which about $8\%$ will be captured instead
of decaying~\cite{lsndNIM}.  Almost all of these captures are on
$^{12}$C nuclei, rather than free protons, and all but about $16\%$
will go to particle-unbound excited states of
$^{12}$B~\cite{mucapture}.  Note that low-energy protons and alpha
particles will be invisible in MiniBooNE because of the low
scintillation yield and the effects of light quenching.  The rate of
captures to the ground state of $^{12}$B is thus about 11 Hz.  This
isotope is unstable to $\beta^-$ decay, with mean lifetime 20 ms and
electron total energy endpoint 13.9 MeV.  The shape of the electron
total energy spectrum is
\begin{equation}
\frac{dN}{dE_e} \propto (13.9 - E_e)^2 \, E_e \sqrt{E_e^2 - m_e^2}\,,
\end{equation}
where all energies are in MeV and the normalization is set by the rate
of 11 Hz.  We have neglected the Fermi function, since it causes very
little distortion at these high electron energies.  With the above
considerations for the trigger design, the $^{12}$B lifetime is so
long that a holdoff time cannot be used.  However, most of the
$^{12}$B beta decays will produce events well below the typical
supernova event energies (about $80\%$ of the $^{12}$B beta-decay
electrons have energies below 10 MeV).

The LSND collaboration observed about twice as many low-energy events
that appeared to be $^{12}$B beta decay as expected~\cite{lsndNIM}.
These events were identified by their energy, not their lifetime, so
other muon-induced radioactivities could also contribute.  The origin
of this discrepancy is unknown, but will be investigated further in
MiniBooNE.


\subsection{Other Backgrounds}

At energies below about 5 MeV, the background rates from a wide
variety of radioactive contaminants will rise very quickly.  These
events do not overlap our supernova signal region, and can easily be
removed at the trigger level by requiring a minimum number of hit
phototubes.  The possibility of large backgrounds not considered here
can be excluded empirically by the results from the LSND
detector~\cite{lsndNIM}, which was also located at very shallow depth
and used a similar trigger.


\begin{figure}[t]
\centerline{\epsfxsize=3.25in \epsfbox{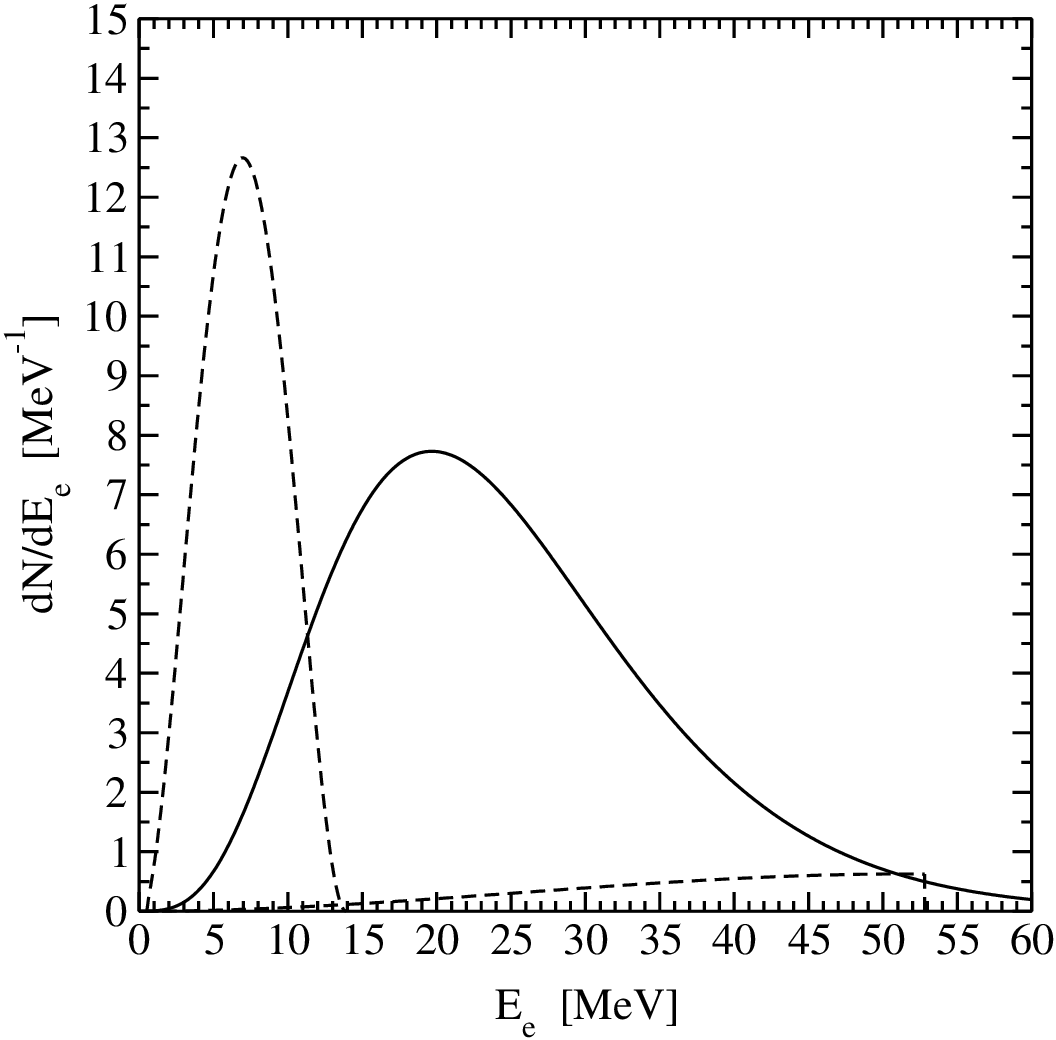}}
\caption{Spectra of the supernova signal (solid line), $^{12}$B decay
background (dashed line peaking at low energy), and surviving muon
decay background (dashed line peaking at high energy) versus the true
electron total energy, over a 10 s interval assumed to contain the
full supernova signal.  A volume of 0.595 ktons is assumed, though all
rates are reduced by $15\%$ to account for the detector deadtime
fraction imposed by applying a $15.2\ \mu$s holdoff after any muon
event.  Energy resolution is not included.  Below about 5 MeV,
backgrounds from ambient radioactivities will dominate over the
spectra shown.}
\label{fig:spectra1}
\end{figure}

\begin{figure}[t]
\centerline{\epsfxsize=3.25in \epsfbox{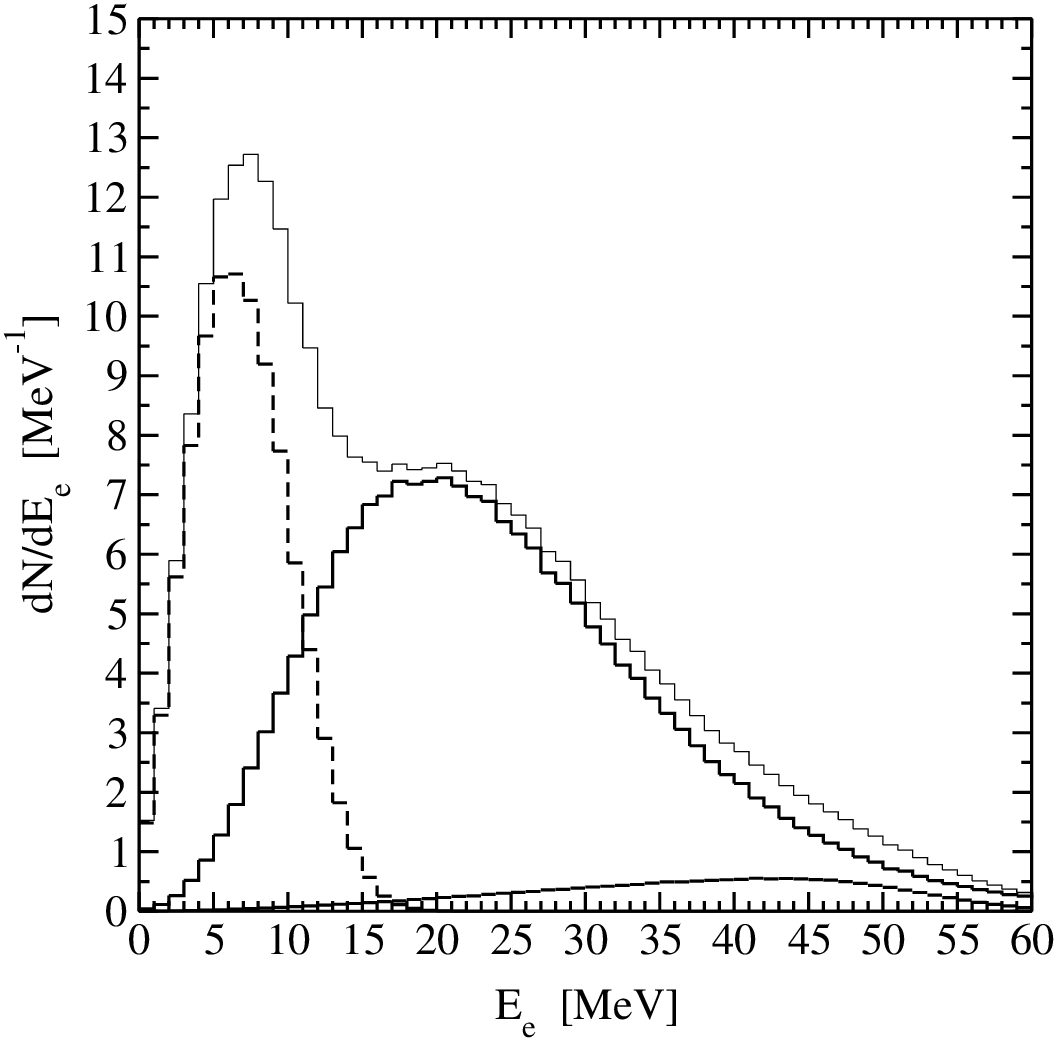}}
\caption{The same as Fig.~\ref{fig:spectra1}, except that energy
resolution is now implemented as described in the text.  The thin
solid line indicates the sum spectrum.  The curves shown indicate the
true spectral shapes.  For an actual supernova, there will be Poisson
fluctuations on the numbers of events shown in each of the (1 MeV
wide) bins.}
\label{fig:spectra2}
\end{figure}

\section{Results}

In Fig.~\ref{fig:spectra1} we show the theoretical shapes of the
supernova neutrino events, as well as the $^{12}$B decay and surviving
muon decay backgrounds, over a 10 s interval assumed to contain the
full supernova signal.  All three classes of events are nearly
isotropic, and will be nearly uniformly distributed in position.

The supernova signal is well above most radioactive backgrounds in
energy, and reasonably above that from $^{12}$B.  It is also well
below the large energy depositions from muons.  Michel decays from
stopped muons do lie in the same energy range as supernova neutrino
events, and their rate is about 2 kHz.  However, we have shown that
these background events can easily be reduced to a rate of about 2 Hz.

In Fig.~\ref{fig:spectra2} we have taken the estimated energy
resolution (see above) of the detector into account.  It is shown that
this has a relatively minor effect on the spectra.

We have assumed that muons can be identified with very high efficiency
by requiring either a large number of hit phototubes in the veto
region {\bf OR} the main detector volume.  We can then impose a $15.2\
\mu$s holdoff after any such event.  This is over-conservative in the
sense that most of these muons will not actually stop and decay in the
detector, but the penalty is minor, just a $15\%$ deadtime.  With a
modest cut at low energies, i.e., requiring a minimum number of hit
phototubes, the low-energy radioactivities and a good deal of the
$^{12}$B beta decays can be cut.  In sum, the steady-state rate should
be about 4 Hz, easily manageable by the data acquisition electronics.

\begin{figure}[t]
\centerline{\epsfxsize=3.25in \epsfbox{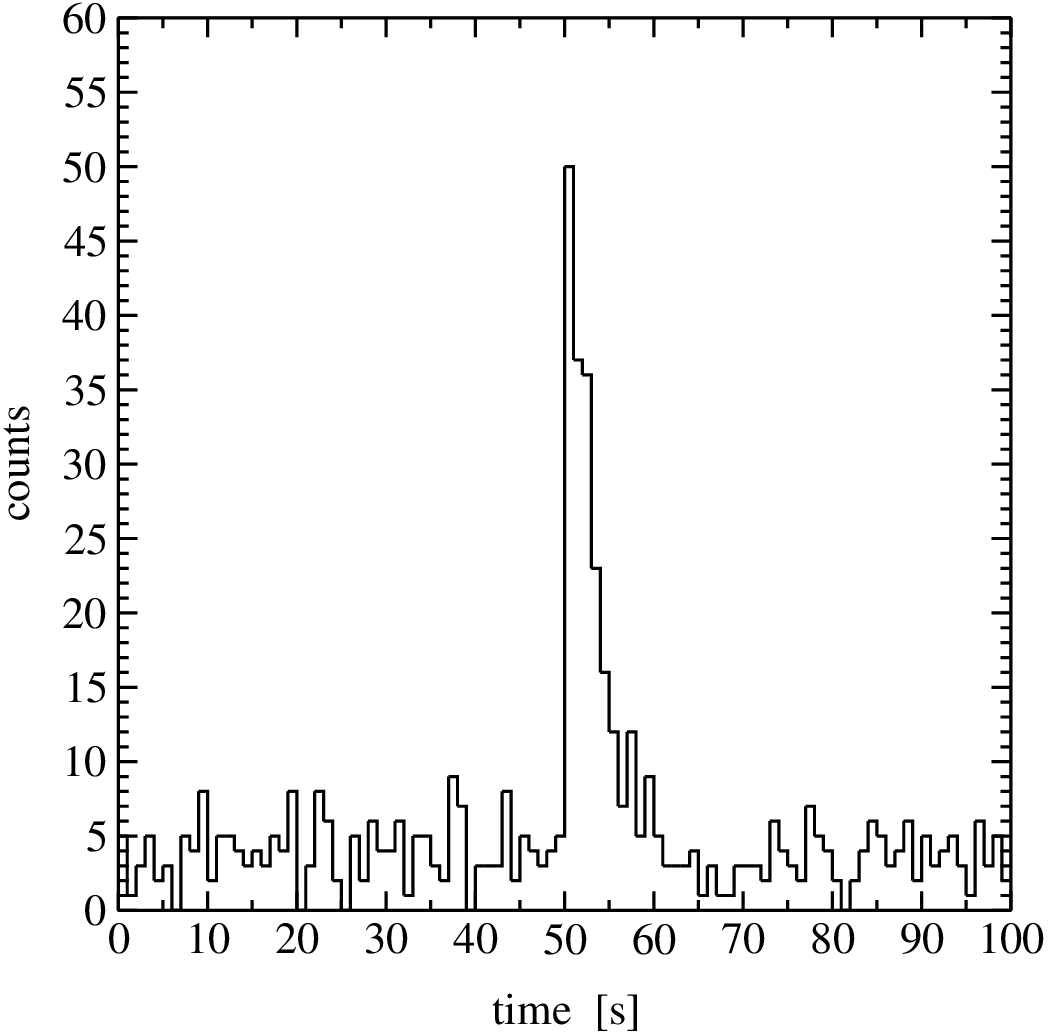}}
\caption{Total number of signal and background events, in 1 s bins
over 100 s, showing the Poisson fluctuations for one random simulated
experiment with a supernova at $t = 50$ s.  The background rates above
10 MeV from $^{12}$B beta decays and muon decays are each about 2 Hz.
All rates have been multiplied by 0.85 to account for detector
deadtime, as discussed in the text.  The supernova signal was modeled
as a sharp rise followed by an exponential decay with time constant 3
s.}
\label{fig:burst}
\end{figure}

A candidate supernova can be flagged by a large increase in the data
rate, as shown in Fig.~\ref{fig:burst}.  A circular buffer can store
data for offline evaluation, where it can be examined to see if it has
reasonable characteristics (energy spectrum, duration, event positions
and directions, etc.).  Detailed discussions of supernova trigger for
offline evaluation systems were published for
Kamiokande~\cite{trigkam} and MACRO~\cite{trigmacro}.


\section{Discussion and Conclusions}

The MiniBooNE experiment~\cite{miniboone} will decisively test the
neutrino oscillation signal reported by LSND~\cite{lsnd}.  If the
signal is confirmed, it will have a big impact on all of neutrino
physics, since simple models with three active neutrinos appear to be
inadequate to explain all the data.  In addition, several authors have
shown that the required mixing parameters would have interesting
implications for various aspects of core-collapse supernovae,
including the explosion mechanism, $r$-process production of the heavy
elements, and the detected neutrino signal~\cite{SNlsnd}.

Our results show that MiniBooNE could be quite useful as a supernova
neutrino detector, despite being optimized for much higher energies
and being at a shallow depth of only 3 m.  With very simple
trigger-level cuts, the backgrounds associated with the 10 kHz
cosmic-ray muon rate can easily be reduced to a manageable level, as
shown in Figs.~\ref{fig:spectra1} and ~\ref{fig:spectra2}.  The
approximately 230 events from a canonical Galactic supernova at 10 kpc
can thus be easily identified, with only minimal background
contamination.  Only about $15\%$ of these events will be lost to
detector deadtime as a result of cuts to reduce the muon decay
background.  This leaves about 190 supernova events, and their
spectrum should be well-measured.  The steady-state data rate of about
4 Hz in the data acquisition electronics is also easy to handle.  The
details of implementing a supernova trigger into the MiniBooNE data
acquisition system are now being studied.  Further, in the very near
future, direct measurements of the detector performance and
backgrounds will be measured in detail.

What can MiniBooNE add to the worldwide effort to detect supernova
neutrinos?  {\it First}, it is highly desirable to have as many
different detectors as possible.  This will allow important cross
checks of the results, both from a theoretical and an experimental
point of view.  {\it Second}, MiniBooNE may be able to act as a node
in the Supernova Early Warning System (SNEWS)~\cite{SNEWS}.  While
triangulation of the supernova direction by arrival-time differences
in several detectors likely remains very difficult~\cite{pointing},
having many independent nodes in the network greatly reduces the false
alarm rate.  Also, since neutrinos leave the proto-neutron star hours
before light leaves the stellar envelope, detection of supernova
neutrinos may allow for astronomical observations of the earliest
stages of the supernova.  {\it Third}, not all detectors are live all
the time, due to upgrades and calibrations.  Until SK is repaired, the
$\bar{\nu}_e + p \rightarrow e^+ + n$ yield in MiniBooNE would be
comparable to that from other detectors with hydrogen targets.  {\it
Fourth}, the signal in MiniBooNE may be useful for studying
matter-affected mixing effects on neutrino propagation in Earth,
especially when compared to other $\bar{\nu}_e + p \rightarrow e^+ +
n$ detectors at different locations.  These matter effects can
significantly distort the spectrum of detected positrons (see, e.g.,
Ref.~\cite{lunardini}).

We have shown that MiniBooNE can function as a useful supernova
neutrino detector, despite its high cosmic-ray related background
rates.  One immediate application of this technique is that other
surface-level neutrino detectors may be also be useful for detecting
supernova neutrinos.


\section*{ACKNOWLEDGMENTS}

We thank Andrew Bazarko, Len Bugel, Eric Church, Bonnie Fleming, Gerry
Garvey, Richard Imlay, Jon Link, Bill Metcalf, Peter Meyers, Jen Raaf,
Richard Schirato, Mike Shaevitz, Michel Sorel, Rex Tayloe, and
especially Steve Brice, Janet Conrad, Bill Louis, and Geoff Mills for
very valuable discussions.

M.K.S. was supported by the Rabi Undergraduate Scholars Program of
Columbia University.  M.K.S. and J.A.F. were additionally supported by
NSF grant PHY-0098826.  J.F.B (as the David N. Schramm Fellow) was
supported by Fermilab, which is operated by URA under DOE contract
No.\ DE-AC02-76CH03000.  J.F.B. was additionally supported by NASA
under NAG5-10842.



\end{document}